# Integration of 5G Technologies in LEO Mega-Constellations

Alessandro Guidotti[1], Alessandro Vanelli-Coralli[1], Oltjon Kodheli[1], Giulio Colavolpe[2], Tommaso Foggi[2]

[1] Dept. of Electrical, Electronic, and Information Engineering (DEI), Univ. of Bologna, Bologna
Email: {a.guidotti, alessandro.vanelli}@unibo.it, oltjonkodheli@gmail.com

[2] Dept. of Information Engineering, Univ. of Parma, Parma, Italy
Email: giulio.colavolpe@unipr.it, tommaso.foggi@gmail.com

**Abstract:** 3GPP is finalising the first release of the 5G New Radio physical layer. To cope with the demanding 5G requirements on global connectivity and large throughput, Satellite Communications might be a valuable resource to extend and complement terrestrial networks. In this context, we introduce an integrated architecture for 5G-based LEO mega-constellations and assess the impact of large Doppler shifts and delays on both the 5G waveform and the PHY/MAC layer procedures.

**Introduction**

Satellite Communication (SatCom) systems, thanks to their inherently large footprint, provide a valuable and cost-effective solution to complement and extend terrestrial networks, not only in rural areas and mission critical situations, but also for traffic off-loading in densely populated areas. In this context, the integration of Long Term Evolution (LTE) in Low Earth Orbit (LEO) mega-constellation systems, *i.e.*, hundreds of satellites, is gaining an ever increasing attention, [1]-[9], as also demonstrated by several recent commercial endeavors. Furthermore, 3GPP Radio Access Network (RAN) activities are now entering in a critical phase, with the first physical layer (PHY) standard for 5G systems almost completed, and this provides a unique chance to define a fully-fledged satellite-terrestrial architecture. In particular, in [8]-[9], 3GPP started the activities related to Non-Terrestrial Networks, a new study item in RAN activities. In this article, after introducing a candidate architecture for the integration of 5G technologies and procedures in a LEO mega-constellation SatCom system, we focus on the related technical challenges by outlining the criticalities introduced by typical satellite channel impairments, *i.e.*, large Doppler shift and propagation delays, on both the waveforms and the PHY/MAC layer procedures, which are characterized by strict timing requirements.

**System Architecture and Problem Statement**

We consider a LEO mega-constellation system providing backhaul connectivity to several satellite-enabled on-ground network entities. In this system, shown in Figure 1, the terrestrial 5G User Equipments (UEs), denoted as New Radio (NR), in each on-ground cell are connected to the satellite-enabled entity through a terrestrial 5G link. The satellites in the LEO mega-constellation are assumed to be transparent and the gateway is connected to the satellite through an ideal feeder link providing access to the 5G Core Network (CN). Please note that the use of regenerative satellites is also currently being discussed in 3GPP RAN meetings, in addition to the transparent case, [8]. However, if not otherwise specified, in the following we exclusively refer to transparent satellites. Finally, Frequency Division Duplexing (FDD) is assumed for the frame structure, since Time Division Duplexing is not feasible due to the large delays.

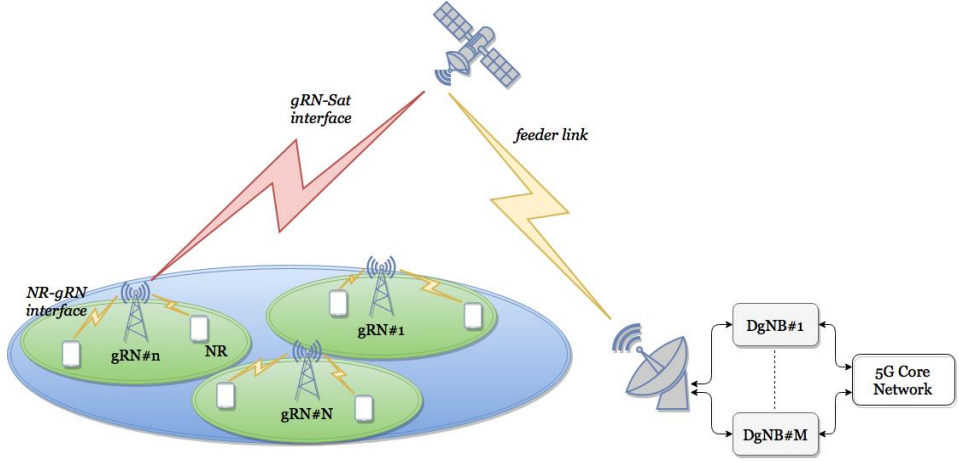

Figure 1 - Sat-gRN system architecture.

We assume the satellite-enabled network entities to be Relay Nodes (RNs), *i.e.*, low-power base stations connected to a Donor gNB (DgNB), where the gNB is the 5G evolution of the LTE eNB. This architecture is referred to as Sat-gRN. RNs are defined in LTE, [10], while not yet for the future 5G architecture. However, it is agreed that they will be implemented in future releases, [11]. Thus, in the following, we assume that 5G RNs, referred to as gRNs, follow the same principle of those standardised in LTE. In particular, the same Air Interface between the gRN and the LEO satellites as that between the traditional NRs and gNBs is considered and the gRN terminates the radio protocols up to Layer 3. Based on these observations, the gRN is seen as a gNB by the NR and as a NR by the corresponding DgNB. In the proposed system architecture, the DgNBs are conceptually located at the system gateway and interact with the 5G CN. It is worthwhile highlighting that each gRN creates a normal terrestrial 5G cell to which the NR connects trough a normal NR Air Interface, *i.e.*, the terrestrial access link requires no modifications in the proposed system. Thus, the impact of large delays and Doppler shifts will be assessed referring to the gRN-Sat Air Interface.

Propagation Delay: To compute the one-way propagation delay, *i.e.*, gRN-Sat-GW link, similarly to one of the non-terrestrial network scenarios discussed in [8] for single-satellite deployments, we consider LEO satellites at an altitude $h = 1500$ km. The minimum distance between them and the gRN is reached with a 90° elevation angle, while the maximum distance depends on the elevation angle. In particular, we assume a $\vartheta_{gRN} = 10°$ for the gRN-Sat elevation angle and $\vartheta_{GW} = 5°$ for the GW-Sat one. The one-way propagation delay in the worst-case scenario is thus given by, [8]:

$$T_{prop-1way} = T_{gRN-Sat} + T_{Sat-GW} = 12.158 + 13.672 = 25.83\ ms \qquad (1)$$

The two-way propagation delay is twice this value:

$$T_{prop-2way} = 2T_{prop-1way} = 51.66\ ms \qquad (2)$$

This value, as discussed in the following, is significantly larger than the maximum timing requirements allowed by 4G and 5G PHY/MAC procedures and, thus, its impact shall be properly assessed to understand whether the same 5G technologies and algorithms can be implemented or if proper modifications are needed in the considered scenario. This is true, in particular, without even considering the processing time, which can be assumed to be equal to that in LTE networks and, thus, negligible (typically, 1-2 ms) with respect to the large propagation delays over satellite links showed in (1)-(2). It shall be highlighted that this is a worst-case scenario, when the UE and the GW are at very low elevation angles as reported in [8]. However, when we deploy a LEO mega-constellation, it is unlikely that the serving satellite will be the one almost on the horizon, due to the very large number of available satellites. As discussed in [1]-[2], the lowest elevation angle between the serving satellite and the gRN is likely to be approximately 45°. In any case, the analysis reported in this paper is still applicable and it provides results for a worst-case scenario.

Doppler: The target user mobility in 5G requirements is set to 500 km/h for frequencies below 6 GHz and is defined as the maximum NR speed with respect to the serving gNB at which the NR can be served with a predefined QoS. Considering a 4 GHz carrier frequency, the maximum Doppler shift is easily found to be 1.9 kHz. When considering the gRN-Sat link, the Doppler shift is caused exclusively by the satellite movement on its orbit. In particular, assuming a system in Ka-band between 20 GHz (DL) and 30 GHz (UL) at $h = 1500$ km, we have $v_{sat} = 7.1171$ km/s and a maximum Doppler shift in the range $400 \text{ kHz} < f_d < 600 \text{ kHz}$, [8]. It can be noticed that these values are far above the maximum Doppler shifts foreseen in terrestrial 5G links and it might significantly impact the NR waveform.

**Technical Challenges and Possible Solutions**
Waveform: 3GPP already agreed to rely on CP-OFDM at least for enhanced Mobile Broadband (eMBB) and ultra Reliable and Low Latency Communications (uRLLC) services, coupled with a flexible and scalable numerology, for both uplink and downlink transmissions. However, it shall be noted that, in the uplink, the legacy LTE SC-FDMA waveform can still be implemented. The subcarrier spacing (SCS) reported in 3GPP specifications are defined as multiples of the LTE traditional spacing as $SCS = 15 \cdot 2^n$ kHz, where $n$ is a non-negative integer value. The previously computed Doppler shift experienced by gRNs in the considered scenario is significantly above the foreseen SCS for NR waveforms and suitable solutions are thus required.

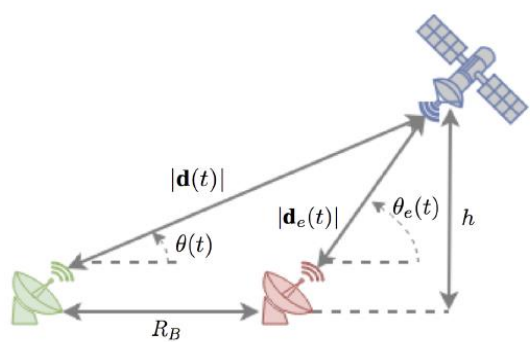

*Figure 2 - gRN position estimation.*

The Doppler shift can be pre-compensated at the system gateway by equipping the gRN with GNSS receivers providing the gRNs estimated positions. This information, together with the satellite orbits, can be exploited at the gateway to pre-compensate for the Doppler shift to a high extent. From Figure 2, the distance between the gRN and the satellite, $|d(t)|$, and the elevation angle, $\vartheta(t)$, can be computed through a simple geometry analysis, [1]-[2]. When considering an estimation error on the gRN location, which leads to a wrong elevation angle $\vartheta_e(t)$ and distance to the satellite $|d_e(t)|$, a residual Doppler shift will occur and this value shall be limited to be within the maximum Doppler shift that can be afforded by the NR SCS, which is approximately 30.4 kHz when $SCS = 480$ kHz.

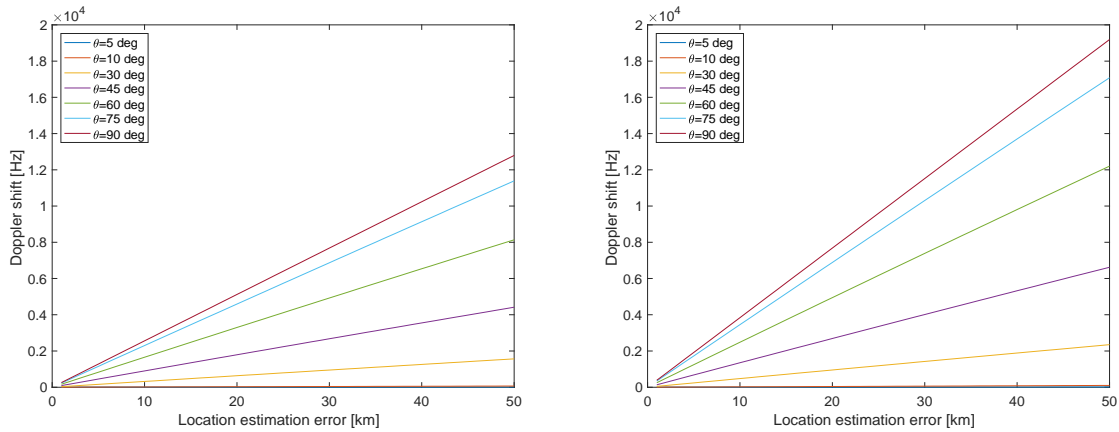

*Figure 3 - Residual Doppler shift as a function of the elevation angle and the location estimation error for the DL (left, 20 GHz) and the UL (right, 30 GHz).*

By means of geometrical considerations based on Figure 2, the residual Doppler shift can be computed as a function of the estimation error and the satellite elevation angle, as shown in Figure 3 for different elevation angles, thanks to the approach proposed in [1]-[2]. It shall be noted that the largest residual Doppler shift is experienced with a 90° elevation angle, while at the lowest considered elevation angles it is negligible.

With respect to the PHY/MAC procedures, 3GPP activities are still ongoing. However, the main characteristics of these procedures have been defined, while some detailed parameter descriptions are still to be completed. In order to analyse the impact of large RTT and Doppler shifts on these procedures, when no parameters are defined for NRs, data from the LTE standards will be considered.

Random Access: 3GPP defined that the NR RA procedure will be similar to that in LTE, *i.e.*, either contention-based or contention-free, depending on whether it is performing a simple handover or connecting for the first time to the 5G network, respectively. Focusing on the contention-based procedure, since the contention-free approach corresponds to its first two steps, it can be noticed that two different timers are present, [1]-[2], [12]: the Random Access Response (RAR) time window in Step 2, which can be set up to 15 ms, and the contention resolution timer in Step 4, which can be as large as 64 ms. With respect to the previously defined maximum two-way propagation delay $T_{prop-2way}$ in the proposed system equal to 51 ms, while the contention resolution timer is above this value and poses no challenges, the RAR time window is lower than the expected communication delay. On one hand, a possible solution would be to increase the maximum RAR timer to a value larger than the 51 ms delay. On the other hand, we can notice that the RA procedure comes into play on the gRN-Sat Air Interface only during the gRN start-up, *i.e.*, during the first phase of the gRN attach procedure. Once the gRN is connected to its DgNB, no further RA procedures are required since this is a fixed network element. By observing that both the satellites orbits and the gRN locations are known, this technical challenge can be circumvented by implementing an ad hoc network deployment procedure at gRN start-up.

HARQ: The HARQ protocol foreseen by 3GPP for 5G systems is based on the LTE Stop-and-Wait (SAW) parallel HARQ processes. The minimum number of parallel HARQ processes is defined as the ratio between the HARQ processing time, *i.e.*, ACK time window plus propagation delay, and the Transmission Time Interval (TTI), *i.e.*, $T_{HARQ}/TTI$. It is agreed that NRs should support multiple HARQ configurations to enhance the system flexibility, while in LTE only a configuration with 8 parallel HARQ processes is allowed. In the considered scenario, the critical parameter is the 51 ms propagation delay, which impacts $T_{HARQ}$. In particular, assuming a 1 ms TTI and an 8 ms time window for ACK reception as in LTE, we can find that the minimum number of parallel HARQ processes is $N_{HARQ} = (51 + 8)/1 = 59$. This large number of parallel processes impacts the NR soft-buffer size, which shall be proportional to $N_{HARQ} \cdot TTI$, and the bit-width of Downlink Control Information (DCI) fields (3 in LTE because there are 8 processes).

To cope with the above issue, several solutions can be envisaged: i) increasing the buffer size to cope with the large number of HARQ processes; ii) increase the number of HARQ processes, by maintaining the buffer size under control, by using a 2 bit ACK, [13], to inform the transmitter on how close the received packet is to the originally transmitted one. Therefore, the number of retransmission will be reduced, because the transmitter can add the redundant bits according to the feedback information; iii) reducing the number of HARQ processes and the buffer size, which also reduces the system throughput; and iv) not implementing the HARQ protocol, which requires solutions to solve issues related to colliding/non-decodable packets.

**Conclusion**

In this article, we introduced a possible architecture for the integration of 5G technologies and LEO mega-constellations, focusing on the main technical challenges related to both 5G waveforms and PHY/MAC procedures, based on the latest 3GPP specifications. The impact of the Doppler shift on the waveform can be compensated by accurate GNSS receivers, also considering the increased subcarrier spacings available for NRs. The impact of large propagation delays on the RA procedure can be limited by either adopting ad hoc network deployment at the gRN start-up or by increasing the RAR timer. Finally, we derived that at least 59 HARQ parallel processes are needed in the considered scenario, posing a significant burden on the NR buffer size and DCI field. Some solutions have been proposed for keeping the number of HARQ processes and buffer size under control. Finally, it shall be noted that we considered transparent satellites. In case regenerative satellites are assumed, a different analysis is required. In particular, the two-way propagation delay only involves the gRN-Sat link in this case and, thus, we would have $T_{prop-2way} = 24.32$ ms, [8], which is half the case of transparent satellites.

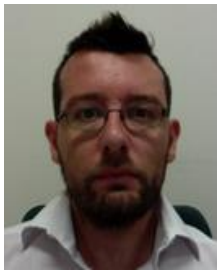

**Alessandro Guidotti** (Member, IEEE) received the Master degree (magna cum laude) in Telecommunications Engineering and the Ph.D. degree in Electronics, Computer Science, and Telecommunications from the University of Bologna in 2008 and 2012, respectively. He is Research Associate in the Department of Electrical, Electronic, and Information Engineering (DEI) "Guglielmo Marconi" at the University of Bologna. His research interests include wireless communication systems, spectrum management, cognitive radios, interference management, and stochastic geometry.

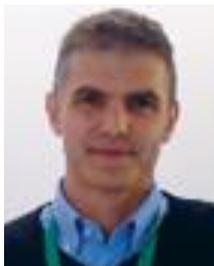

**Alessandro Vanelli-Coralli** (Senior Member, IEEE) received the Dr. Ing. Degree in Electronics Engineering and the Ph.D. in Electronics and Computer Science from the University of Bologna (Italy) in 1991 and 1996, respectively. In 1996, he joined the University of Bologna, where he is currently an Associate Professor and Chair of the PhD Board in Electronics, Telecommunications and Information Technologies. Dr. Vanelli-Coralli participates in national and international research projects on satellite mobile communication systems: he has been the Project Coordinator of the FP7 STREP CoRaSat (Cognitive Radio for SatCom), and Scientific Responsible for several European Space Agency and European Commission funded projects. He is an IEEE Senior Member.

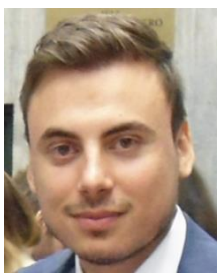

**Oltjon Kodheli** received the Master Degree in Electronic Engineering (cum laude) from University of Bologna (Italy) in October 2016 and the Bachelor Degree in Electronic Engineering from Polytechnic University of Tirana (Albania) in 2013. He currently holds a grant at University of Bologna for performing research related to 5G Radio Access Network. His research activities are mainly focused on wireless communication systems and multicarrier modulation techniques.

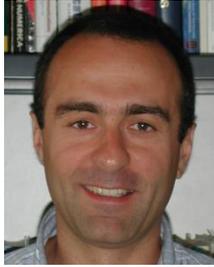

**Giulio Colavolpe** (Senior Member, IEEE) (S'96-M'00-SM'11) received the Dr. Ing. degree in Telecommunications Engineering (cum laude) from the University of Pisa, Italy, in 1994 and the Ph.D. degree in Information Technologies from the University of Parma, Italy, in 1998. Since 1997, he has been at the University of Parma, Italy, where he is now Professor of Telecommunications at the Dipartimento di Ingegneria e Architettura (DIA His research interests include the design of digital communication systems, adaptive signal processing (with particular emphasis on iterative detection techniques for channels with memory), channel coding and information theory. His research activity has led to more than 200 papers in refereed journals and in leading international conferences, and 18 industrial patents.

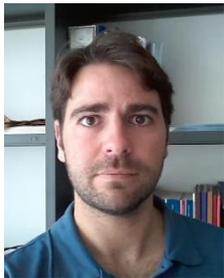

**Tommaso Foggi** received the master's degree in Telecommunication Engineering from the University of Parma in 2003 and the Ph.D. degree in Information Technology from the same University in 2008. Since 2009, he is a research engineer of National Inter-University Consortium for Telecommunications (CNIT) in the University of Parma research unit. His main research interests include electronic signal processing for optical communication systems, in particular adaptive equalization, coding and iterative decoding techniques, optical channel impairment compensation, channel estimation, simulative software implementation.